\documentclass[showpacs,
floatfix,
aps,
prb,
amsmath,
twocolumn,
superscriptaddress,
tightenlines]{revtex4-1}
\usepackage{graphicx}
\usepackage{dcolumn}
\usepackage{amsmath}
\usepackage{amsfonts}
\usepackage{bm}
\usepackage{color}

\newcommand{\be}{\begin{equation}}
\newcommand{\ee}{\end{equation}}
\newcommand{\bea}{\begin{eqnarray}}
\newcommand{\eea}{\end{eqnarray}}

\renewcommand{\vec}[1]{{\bm #1}}

\def\E{\mathbf{E}}
\def\R{\mathbf{R}}
\def\Edc{\mathcal{E}_H}
\def\Eac{E}
\def\F{\mathcal{F}}

\def\R{\mathcal{R}}

\def\tst{\tau_\star}
\def\eac{\epsilon}
\def\edc{\epsilon_{\rm dc}}

\def\oc{\omega_{\mbox{\scriptsize {c}}}}

\def\rc{R_{\mbox{\scriptsize {c}}}}

\def\tq{\tau_{\mbox{\scriptsize {q}}}}

\def\ttr{\tau_{\mbox{\scriptsize {tr}}}}

\def\tsh{\tau_{\mbox{\scriptsize {sh}}}}
\def\tsm{\tau_{\mbox{\scriptsize {sm}}}}
\def\ttr{\tau}

\def\stdir{\mathrm{St}_{\mathrm{dir}}}
\def\stint{\mathrm{St}_{\theta}}
\def\stdc{\mathrm{St}_{\mathrm{dc}}}

\def\ey{\hat{\mathbf e}_y}
\def\ex{\hat{\mathbf e}_x}
\def\e{\hat{\mathbf e}}
\def\n{{\hat{\mathbf n}}}

\newcommand{\req}[1]{Eq.\,(\ref{#1})}

\newcommand{\rfig}[1]{Fig.\,\ref{#1}}

\newcommand{\rref}[1]{Ref.\,\onlinecite{#1}}

\begin{document}

\title{Phase-sensitive bichromatic photoresistance in a two-dimensional electron gas}

\author{Q.~Shi}
\affiliation{School of Physics and Astronomy, University of Minnesota, Minneapolis, Minnesota 55455, USA}
\author{M.~Khodas}
\affiliation{Department of Physics and Astronomy, University of Iowa, Iowa City, Iowa 52242, USA}
\author{A.~Levchenko}
\affiliation{Department of Physics and Astronomy, Michigan State University, East Lansing, Michigan 48824, USA}
\author{M.~A.~Zudov}
\affiliation{School of Physics and Astronomy, University of Minnesota, Minneapolis, Minnesota 55455, USA}

\begin{abstract}
We have studied microwave photoresistance in a two-dimensional electron system subject to two radiation fields (frequencies $\omega_1$ and $\omega_2$) using quantum kinetic equation. 
We have found that when $\omega_2/\omega_1= 1 + 2/N$, where $N$ is an integer, and both waves have the same polarization, the displacement mechanism gives rise to a new, phase-sensitive photoresistance.
This photoresistance oscillates with the magnetic field and can be a good fraction of the total photoresistance under typical experimental conditions.
The inelastic mechanism, on the other hand, gives zero phase-sensitive photoresistance if the radiation fields are circularly polarized.
\end{abstract}

\received{July 3, 2013}


\maketitle

Two-dimensional electron systems subject to both perpendicular magnetic field and microwave radiation reveal a variety of fascinating nonequilibrium transport phenomena,\citep{dmitriev:2012} including microwave-induced resistance oscillations (MIROs)\citep{zudov:2001a,ye:2001} and zero-resistance states (ZRSs).\citep{mani:2002,zudov:2003,yang:2003}
MIROs have been discussed in terms of the displacement mechanism,\citep{ryzhii:1970,durst:2003,lei:2003,vavilov:2004,dmitriev:2009b} which originates from microwave-assisted impurity scattering off disorder, and the inelastic mechanism,\citep{dmitriev:2005,dmitriev:2009b} which stems from radiation-induced modification of the electron distribution function.
Both mechanisms predict that the photoresistance oscillates as $-\sin (2\pi\omega/\oc)$, $\omega = 2\pi f$ is the microwave frequency, and $\oc$ is the cyclotron frequency.
ZRSs, on the other hand, are explained in terms of current domains resulting from electrical instability of a homogeneous state with negative conductivity.\citep{andreev:2003,auerbach:2005,finkler:2009,dorozhkin:2011}

With few exceptions,\citep{joas:2004,zudov:2006a,zudov:2006b,lei:2006a,lei:2006c,kunold:2007} studies of MIROs and ZRSs were limited to the case of monochromatic excitation.
It is known, however, that excitation by two radiation fields, 
$\E_{i}=\mathrm{Re}\{E_i \mathbf{e}_{i}e^{-i \omega_{i}t}\}$
and
$E_i  = \mathcal{E}_i  e^{-i \theta_i}$ ($i=1,2$),
has important implications in many disciplines, such as chemistry,\citep{azaro:1988,shapiro:2003} semiconductor physics,\citep{baskin:1988,fraser:1999,bhat:2005,wahlstrand:2006,wahlstrand:2011} atomic physics,\citep{anderson:1992,cavalieri:1997,bolovinos:2008} and nonlinear optics in glass fiber.\citep{anderson:1991}
Of special interest is the case where the two frequencies are commensurate, i.e, are related as $\omega_2/\omega_1 = N_2/N_1$ ($N_1$ and $N_2$ are integers), in which case one can introduce a relative phase $\theta=N_1\theta_2-N_2\theta_1$ between $E_1^{N_2}$ and $E_2^{N_1}$.

The unifying theme of bichromatic excitation is controlling physical or chemical processes by $\theta$.
For instance, quantum interference between two distinct paths formed by absorption and/or emission of different photons can modify the transition rate between the two states, making it $\theta$-dependent.
In systems with (without) an inversion symmetry, interference may occur when $N_1$ and $N_2$ are of the same (arbitrary) parity for generic radiation polarization.\citep{note:11}
For $\omega_2/\omega_1 = 3$, interference has been demonstrated in the ionization rate of sodium\citep{cavalieri:1997}, in the transition rate between bound states of mercury,\citep{chen:1990,chen:1990a} and, for $\omega_2/\omega_1 = 2$, in a coherent carrier population control in non centrosymmetric crystals.\citep{fraser:1999,bhat:2005,wahlstrand:2006}

Even in the absence of interference,\citep{note:14} the relative phase between the two radiation components can still be used to control the angular distribution of ionized electrons in atoms,\citep{anderson:1992} excitation of carriers in semiconductors,\citep{baskin:1988,anderson:1991,atanasov:1996} and branching ratios in photodissociation.\citep{azaro:1988}
For instance, when $\omega_2/\omega_1 = 2$, bichromatic excitation generates a dc photocurrent, which is determined by a third rank tensor $\propto \mathcal{E}_{1}^2 \mathcal{E}_{2}$.\citep{baskin:1988,atanasov:1996}
The direction of this photocurrent is given by a $\theta$-dependent vector $\n$ such that the time-average of $[({\bf E}_1+{\bf E}_2)\cdot \n]^3$ is maximal.\citep{note:10}

In this paper we report on a new phase-sensitive contribution to bichromatic microwave photoresistance which appears when microwaves are circularly polarized.
While this contribution is absent in the inelastic mechanism,\citep{note:13} the displacement mechanism leads to a nonzero phase-sensitive term, provided that the two microwave fields have equal polarizations and that their frequencies are related as $\omega_2/\omega_1=N_2/N_1$, where $|N_1-N_2| = 2$.
Similarly to direct contributions responsible for ordinary MIROs, the phase-sensitive term oscillates with the magnetic field and, for the simplest case of $\omega_2/\omega_1=3$, is maximized (minimized) in the vicinity of half-integer (integer) values of $\eac=\omega_1/\oc$.
We estimate that the phase-sensitive term can be a good fraction of the total photoresistance in samples with a sufficient amount of sharp disorder under typical microwave intensities.

\begin{figure}[t]
\includegraphics[width=3.3in]{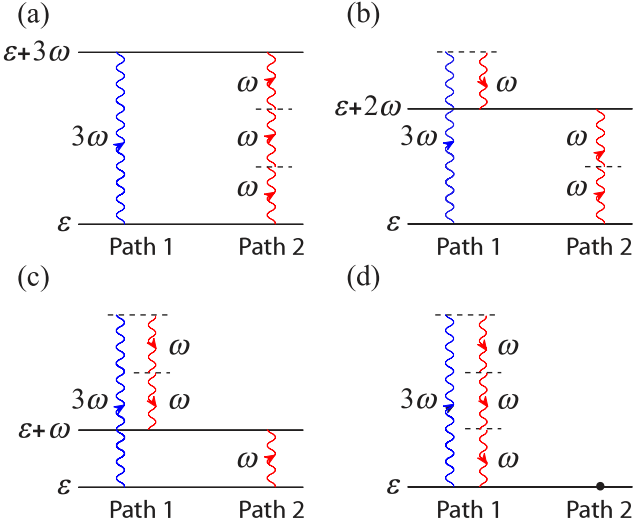}
\vspace{-0.05 in}
\caption{(Color online)
Possible electron transition path pairs under bichromatic ($\omega_1=\omega$ and $\omega_2 = 3\omega$) irradiation corresponding to energy change $\Delta \varepsilon_n =n\hbar\omega$ for (a) $n=3$, (b) $n=2$, (c) $n=1$ and (d) $n=0$. 
For each $n$, one path involves the absorption of one photon of frequency $3\omega$ and emission of $3-n$ photons of frequency $\omega$ and another path is formed by the absorption of $n$ photons of frequency $\omega$.\citep{note:0}
Combined, each path pair involves three photons of frequency $\omega$ and one photon of frequency $3\omega$.
}
\vspace{-0.15 in}
\label{fig1}
\end{figure}

To obtain the conditions necessary for the existence of the phase-sensitive bichromatic photoresistance, we first consider electron scattering off disorder combined with an absorption of a single photon with circular polarization $\sigma=\pm 1$.
It is easy to convince oneself that the amplitude of such a scattering event must contain a phase factor $\exp\left[ i \sigma ( \varphi + \varphi') / 2 \right]$, where $\varphi$ ($\varphi'$) is the angle an electron momentum makes with the dc electric field before (after) the collision with impurity.\citep{note:16}
Indeed, under global rotation ($\varphi \rightarrow \varphi+\alpha$, $\varphi' \rightarrow \varphi'+\alpha$) the above phase factor is multiplied by $\exp( i \sigma \alpha)$, as it should be, to account for the transformation of the circularly polarized field amplitude $E \rightarrow E e^{i\sigma\alpha}$. 

Similarly, the amplitude of a direct $n$-photon absorption process scales with the $n$-th power of the microwave field $\Eac^n$, and its angular dependence contains $\exp\left[ i n\sigma ( \varphi + \varphi') / 2 \right]$.\citep{dmitriev:2007,hatke:2011e}

The above arguments can be extended to processes under bichromatic ($i=1,2$) microwave radiation with frequencies $\omega_{i}=N_{i}\omega$, field amplitudes $\Eac_{i}$, and polarizations $\sigma_{i}$.
A phase-sensitive contribution can be expected when two independent paths exist between two states separated by energy $\Delta\epsilon_{nm} = (m N_1 - n N_2) \omega$, $n = 0,1,...,N_1$, $m=0,1,...,N_2$.
For a given $n$ and $m$, one such path involves absorption of $N_1-n$ photons of frequency $\omega_2$ and emission of $N_2-m$ photons of frequency $\omega_1$ and another path is formed by emission of $n$ photons of frequency $\omega_2$ and absorption of $m$ photons of frequency $\omega_1$.
For $N_1=1$ and $N_2=3$ all four possible path pairs are illustrated in \rfig{fig1}.
It is easy to see that every single path pair requires $N_1$ photons of frequency $\omega_2$ and $N_2$ photons of frequency $\omega_1$. 
As a result, the phase-sensitive contribution to the transition rate, given by the cross term of the two amplitudes, is proportional to $\Eac_2^{N_1-n}(\Eac_1^{N_2-m})^*\times [(\Eac_2^{n})^*\Eac_1^{m}]^* = \Eac_2^{N_1}(\Eac_1^{N_2})^*$.
The angular dependence of this contribution must therefore contain a phase factor $\exp[i (\sigma_2 N_1 - \sigma_1 N_2 ) (\varphi + \varphi')/2]$.
We further note that for odd $|N_1-N_2|$ the corresponding scattering rate [$\propto \Eac_2^{N_1}(\Eac_1^{N_2})^*$] is odd under parity transformation, $\vec{E}_i \rightarrow -\vec{E}_i$, and is therefore forbidden in the absence of the skew scattering.
With the help of these considerations we now make the following conclusions.

First, the phase-sensitive photoresistance is absent in the inelastic mechanism.\citep{note:13}
Indeed, the phase factor $\exp[i (\sigma_2 N_1 - \sigma_1 N_2 ) (\varphi + \varphi')/2]$, entering the scattering amplitude, prevents the modification of the isotropic part of the distribution function; the inelastic photoresistance, proportional to $\langle \exp[i (\sigma_2 N_1 - \sigma_1 N_2 ) (\varphi + \varphi')/2] g_{\varphi - \varphi'} \rangle_{\varphi\varphi'}$ ($g_{\varphi - \varphi'}$ is some function of $\varphi - \varphi'$), vanishes\citep{note:6} for any $\sigma_1$ and $\sigma_2$, as long as $N_1 \neq N_2$.\citep{note:7}

Second, the phase-sensitive photoresistance is nonzero in the displacement mechanism only if 
$|N_1-N_2|=2$ and $\sigma_1=\sigma_2$.
Indeed, for any even $|N_1-N_2|$, the phase-sensitive photoresistance is proportional to $\left < (\sin \varphi - \sin \varphi')^2 \exp[i (\sigma_2 N_1 - \sigma_1 N_2 ) (\varphi + \varphi')/2]\right >_{\varphi\varphi'}$.
Factor $(\sin \varphi - \sin \varphi')^2$\citep{note:8} contains zeroth and second harmonics of $(\varphi + \varphi')/2$, which give rise to nonzero contributions when $|\sigma_2 N_1 - \sigma_1 N_2| =0$
(a trivial case of $N_1=N_2$\citep{note:7}) and when $|\sigma_2 N_1 - \sigma_1 N_2| = 2$, respectively.
We thus conclude that the phase-sensitive displacement contribution is nonzero only when $|N_1-N_2|=2$ and $\sigma_1 = \sigma_2$.\citep{note:15}
In what follows, we employ a quantum kinetic equation to investigate the simplest case of $N_1=1$, $N_2=3$ (the lowest integers when phase-sensitive effects are nonzero).

At large filling factors the dynamics of two-dimensional electrons in a magnetic field is semi-classical.
We consider the experimentally relevant regime of classically strong magnetic fields,
$\oc \ttr \gg 1$, where $\ttr$ is the transport scattering time.
We further assume that the magnetic field is weak enough so that Landau levels are not well resolved, $\oc \tq \lesssim 1$ ($\tq$ is the quantum lifetime).\citep{note:1}
In this case the density of states can be described by
\begin{equation}
\label{dos}
\nu(\varepsilon)=\nu_0 \left[1-2\lambda\cos(2\pi\varepsilon/\oc)\right]\,,
\end{equation} 
where $\lambda=e^{-\pi/\oc\tq}$ is a Dingle factor.  

Like in ordinary MIRO, we are interested in quadratic-in-$\lambda$ contributions, which are largely immune both to thermal broadening of the Fermi surface and to large scale fluctuations of the chemical potential.
The dissipative part of the electric current is determined by the distribution function $f(\varepsilon,\varphi)$,
\begin{equation}\label{j}
j=2ev_F\int d\varepsilon\nu(\varepsilon)\int \frac{d\varphi}{2\pi} \cos\varphi f(\varepsilon,\varphi)\,,
\end{equation}
where $v_F$ is the Fermi velocity, $\varphi$ is the angle the electron momentum makes with the dc electric field, and $f(\varepsilon,\varphi)$ is the solution of the following kinetic equation:
\begin{equation}\label{KE}
\oc\frac{\partial f}{\partial\varphi}\!=
\stdc\{f\}
+
\stdir\{f\}
+
\stint\{f\}
\,.
\end{equation}
Here, the first term, $\stdc\{f\}$, is the collision integral for electron scattering off disorder without participation of microwave quanta. 
The other two terms, $\stdir\{f\}$ and $\stint\{f\}$, account for the modification of the electron-impurity scattering by the bichromatic microwave field $\E_{\mathrm{mw}}=\E_1+\E_2$.
Here we consider the circularly polarized microwaves 
$\E_{i}=\mathrm{Re}\{E_i \e_{i}e^{-i\omega_{i}t}\}$ ($i=1,2$),
$\e_{i}=(\ex+i\sigma_{i}\ey)/\sqrt{2}$, and $\sigma_{i} = \pm 1$ denotes polarization direction. 
The term $\stdir\{f\}$ takes into account only direct (monochromatic) transitions originating from both microwave fields.
These transitions contribute additively to $\stdir\{f\}$ and their rates are proportional to the corresponding intensity.\citep{note:3}
Here, we focus on the phase-sensitive collision term $\stint\{f\}$, which for the case $\omega_1=\omega$, $\omega_2=3\omega$, is given by
\begin{align}\label{KE-BC}
&\stint\{f(\varepsilon,\varphi)\}=
-\frac{\R_1^3 \R_2}{2 \cdot 4!}
\sum_{\pm}\sum_{n=0}^3 {_3C_n}   
\notag \\
& \times \int\frac{d\varphi'}{2\pi}\frac{ [1 - \cos(\varphi - \varphi')]^2}{ \tau_{\varphi - \varphi'} }  \cos \chi_{\varphi\varphi'}
\notag \\
& \times \nu_0^{-1}\nu(\varepsilon\!+\!W_{\varphi\varphi'}\pm n \omega) 
\left[f(\varepsilon\!+\!W_{\varphi\varphi'} \!\pm\! n \omega)\!-\!f(\varepsilon)\right]\, .
\end{align}%
Here, $\pm n \omega$ is the energy change due to microwave absorption (emission) and $_3C_n=3!/(3-n)!n!$.
As illustrated in \rfig{fig1}, for $n=0$ ($n=3$), all three photons of frequency $\omega$ participate in emission (absorption) processes and $_3C_0={_3C_3}=1$. 
However, for $n=2$ ($n=1$) there are three ways of choosing one emitted (one absorbed) photon of frequency $\omega$ and $_3C_1={_3C_2} = 3$.
The factors $\R_i$ account for microwave renormalization of scattering rates $[ 1 - \cos( \varphi - \varphi') ]^2\tau_{\varphi-\varphi'}^{-1}$. 
To the leading order in microwave fields these factors read
\begin{align}\label{R}
\R_{j} = \frac{e v_F \mathcal{E}_i}{ \omega_i (\omega_i + \sigma_i \oc)}\,,\quad i =1,2\,.
\end{align}
The angle $\chi_{\varphi\varphi'}=(3\sigma_1-\sigma_2)(\varphi+\varphi')/2+\theta$ sensitively depends on the phase difference between the two fields, $\theta=\theta_2-3\theta_1$.
The factor $W_{\varphi\varphi'}=e\E_H \cdot \Delta R $ is the energy gained by an electron as a result of a displacement by $\Delta R= \rc \hat{\mathbf z} \times (\n_\varphi-\n_{\varphi'})$ [$\rc = v_F/\oc$ is the cyclotron radius, $\n_\varphi$ ($\n_\varphi'$) is a unit vector of the electron momentum before (after) the collision] in the presence of Hall field, $\E_H$.

\begin{figure}[t]
\includegraphics{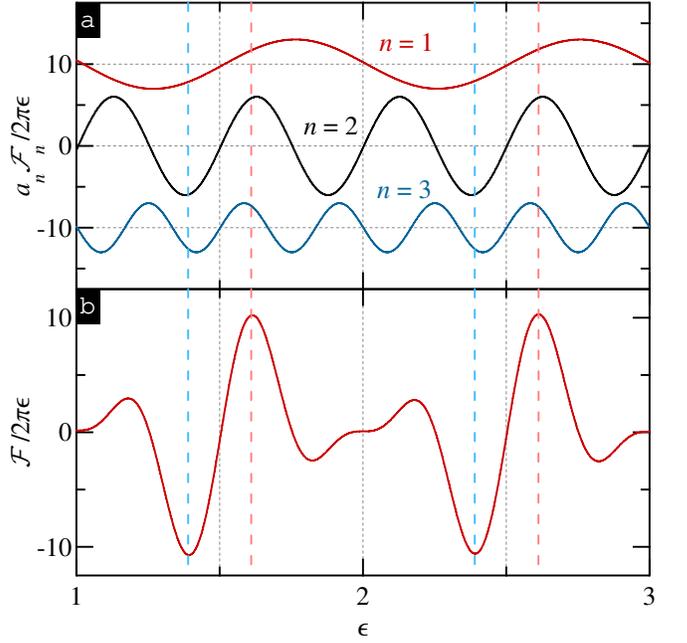}
\vspace{-0.15 in}
\caption{(Color online) (a) $a_n\F_n/2\pi\eac$ versus $\eac$ for $n = 1$ (vertically offset by $+10$), $n=2$, and $n = 3$ (offset by $-10$).
(b) Function $\F/2\pi \eac$ versus $\eac$; see \req{F}.
Extrema are marked by dashed vertical lines, drawn at $\pm 0.11$ from the half-integer $\eac$.
}
\vspace{-0.1 in}
\label{fig2}
\end{figure}

We seek the solution to the kinetic equation, \eqref{KE}, perturbatively in microwave fields, $f(\varepsilon,\varphi)=f_T(\varepsilon)+\delta f(\varepsilon,\varphi)$, where $f_T(\varepsilon)$ is the Fermi distribution function, and 
\begin{equation}\label{delta-f}
\delta f(\varepsilon,\varphi)= A\lambda\partial_\varepsilon
f_T\cos(2\pi\varepsilon/\oc)\cos\varphi\,,
\end{equation}
is the first angular harmonic of the correction to the distribution function resulting from the oscillatory component of $\nu(\varepsilon)$, \req{dos}.
The parameter $A$ is then obtained from \req{KE-BC}, linearized in $\lambda$,
\begin{equation}\label{KE-linear}
-A\lambda \partial_\varepsilon f_T \oc \cos(2\pi\varepsilon/\oc)\sin\varphi=\delta
\stint\{f_T\}+\stint\{\delta
f\}\, .
\end{equation}
By projecting both sides of \req{KE-linear} on $\sin\varphi$ and integrating over angles we obtain 
\begin{equation}\label{A}
A = - \frac{5}{48}
\frac{\edc}{\tilde{\tau}} 
\R_{1}^{3}\R_{2}\delta_{\sigma_1\sigma_2} \F(2\pi\eac)\cos\theta\,,
\end{equation}
where 
$\edc=2e\Edc R_c/\oc$, a parameter which appears in Hall field -induced resistance oscillations.\citep{yang:2002,zhang:2007a,vavilov:2007,hatke:2009c,hatke:2010a,hatke:2011a}
As anticipated, a nonzero $A$ is obtained only for $\sigma_1 = \sigma_2$, so, in what follows, we omit the subscript $\sigma$.
The scattering rate $\tilde{\tau}^{-1}$ entering \req{A} is given by
\begin{align}\label{tau}
\tilde{\tau}^{-1} = \tau_0^{-1} - \frac{3}{2} \tau_1^{-1} + \frac{ 3 }{ 5 } \tau_2^{-1} -\frac{ 1 }{ 10 } \tau_3^{-1} \, ,
\end{align}
where $\tau_k^{-1}$ is the Fourier coefficient of the scattering rate,
$\tau^{-1}_{\varphi-\varphi'} = \sum\limits_k  \tau_k^{-1} e^{ i k (\varphi-\varphi') }$, 
and
\be
\label{F}
\F(x)= \sum\limits_{n=0}^3 a_n \F_n(x)\,,~\F_n(x) = \cos nx -   nx \sin  nx ,
\ee
with the combinatorial factor, $a_n =(-1)^{n+1}{_3C_n}$.

\begin{figure}[t]
\includegraphics{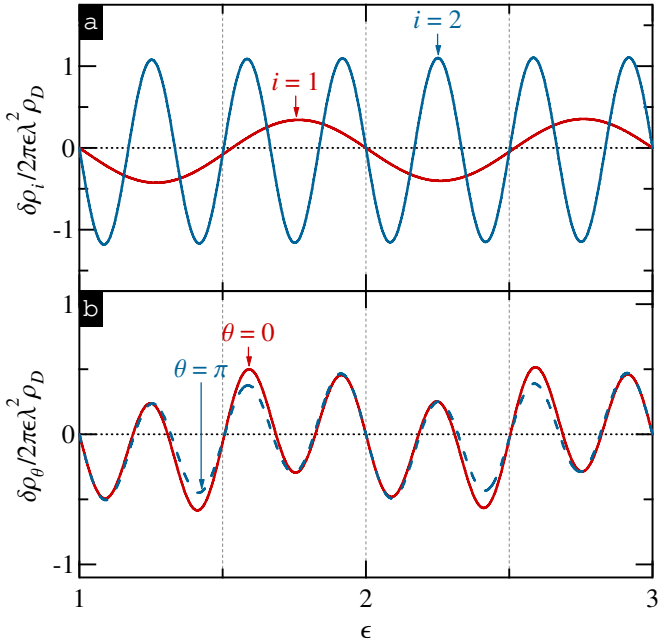}
\vspace{-0.1 in}
\caption{(Color online) Reduced photoresistivity versus $\eac$ (a) due to only direct contributions for $i=1,2$, [\req{direct}], and (b) with the inclusion of the phase-sensitive contribution [\req{rho}] for $\theta = 0$ (solid line) and $\theta =\pi$ (dashed line). 
All curves are drawn for $\R_1=\R_2=0.5$.
}
\label{fig3}
\vspace{-0.15 in}
\end{figure}

Substituting Eqs.\,\eqref{A} and \eqref{delta-f} in \req{j} we obtain the dissipative current and the corresponding phase-sensitive contribution to the photoresistivity,
\begin{equation}\label{rho}
\frac{ \delta\rho_{12}(\theta)}{\rho_D}=\frac{5}{48}
\frac{ \ttr }{ \tilde{\tau} } \lambda^2
\R_1^{3} \R_2
\F(2\pi\eac) \cos\theta\, ,
\end{equation}
where $\rho_D$ is the Drude resistivity.
This equation constitutes the main result of our paper. 

To examine the dependence of the phase-sensitive contribution on $\eac$ we first note that the function $\F(2\pi\eac)$, entering \req{rho}, contains three oscillating terms, with periods of $1$, $1/2$, and $1/3$.
The relative importance of these terms, corresponding to $n = 1$, $2$, and $3$ in \req{F}, respectively, is illustrated in \rfig{fig2}(a), showing $a_k\F_k/2\pi\eac$ as a function of $\eac$.\citep{note:9}
It is easy to see that the derivatives of both $\F_1$ and $\F_3$ have the same (opposite) sign compared to that of $\F_2$ at half-integer (integer) values of $\eac$.
As a result, the sum of these terms is maximized (minimized) in the close vicinity of the half-integer (integer) $\eac$.
Indeed, as illustrated in \rfig{fig2}(b), the largest phase-sensitive contribution occurs at $\eac = k + 1/2 \pm \delta$, where $k=1,2,3,...$ and $\delta \approx 0.11$ [cf. vertical dashed lines].

The total photoresistivity $\delta \rho_{\theta}$ is the sum of the phase-sensitive term, \eqref{rho}, and direct contributions, given by
\begin{align}\label{direct}
\frac{ \delta \rho_i }{\rho_D}
= 
  \frac{ \ttr }{2 \tst } \lambda^2 \R_{i}^2 \left[ \F_1 ( 2\pi \omega_i/\oc) - 1 \right]\,,
\end{align}
where ${\tst}^{-1}=3 \tau_0 ^{-1}-4 \tau_1^{-1}+ \tau_2^{-1}$.

Within the model of mixed disorder, relevant for a high mobility two-dimensional electron system,\citep{vavilov:2007} the Fourier coefficients of the scattering rate are given by
\be
\label{taun}
\tau_k^{-1} = \delta_{k,0}\tsh^{-1} + ( 1 + k^2 \chi)^{-1}\tsm^{-1}\,,
\ee
where $\tsh^{-1}$ and $\tsm^{-1}$ are sharp and smooth disorder scattering rates, respectively, and the parameter $\sqrt{\chi}\ll 1$ is the characteristic scattering angle off the smooth disorder.
The scattering rate $\tilde{\tau}^{-1}$, \req{tau}, then becomes
\be
\label{tau_mixed}
\tilde{\tau}^{-1} = \tsh^{-1} + 36 \chi^3  \tsm^{-1} \, .
\ee
As we see, the contribution of the smooth disorder is weakened because of the presence of a small factor $\chi^3$ in the second term of \req{tau_mixed}.
This suppression is a hallmark of a contribution due to multiphoton processes.\citep{hatke:2011e} 
We therefore limit our remaining discussion to the case of sharp disorder, i.e. $\ttr/\tst = 3$, $\ttr/\tilde\tau = 1$.

We now examine the importance of the phase-sensitive term, \req{rho}, in relation to direct contributions, \req{direct}, which are shown in \rfig{fig3}(a) for both $i=1$ and $i=2$ for $\R_1=\R_2 = 0.5$.
The total microwave photoresistance, which contains both direct and phase-sensitive terms, $\delta \rho(\theta)=\delta\rho_1+\delta\rho_2+\delta\rho_{12}(\theta)$, is presented in \rfig{fig3}(b) for both $\theta = 0$ (solid line) and $\theta = \pi$ (dashed line).
As expected, the difference between the two curves is maximal at the extrema of the phase-sensitive contribution occurring near $\eac = k + 1/2 \pm 0.11$, [see \rfig{fig2}(b)]. 
Quantitatively, we find that near these extrema the relative phase-sensitive contribution, $[\rho(0)-\rho(\pi)]/[\rho_1+\rho_2]$, is quite significant, exceeding 25 \%.
Therefore, the phase sensitivity of the displacement mechanism can be exploited experimentally, for example, to distinguish between the displacement and the inelastic mechanisms.

In summary, we predict a new phase-sensitive contribution to bichromatic microwave photoresistance when microwaves are circularly polarized.
While this contribution is absent in the inelastic mechanism, it is finite in the displacement mechanism, provided that the two microwave fields have equal polarizations and their frequencies are related as $\omega_2/\omega_1=N_2/N_1$, where $|N_1-N_2| = 2$.
Similarly to direct contributions responsible for ordinary MIRO, the phase-sensitive term oscillates with the magnetic field and, for the simplest case of $\omega_2/\omega_1=3$, is maximized (minimized) in the vicinity of half-integer (integer) values of $\omega_1/\oc$.
We estimate that the phase-sensitive term can be a good fraction of the total photoresistance in samples with a sufficient amount of sharp disorder under typical experimental conditions.
Experimental detection of this photoresistance effect will be the subject of future studies.

We thank I. Dmitriev, I. Gornyi, and B. Shklovskii for critical comments and useful discussions.
The work at University of Minnesota was supported by the U.S. Department of Energy, Office of Basic Energy Sciences, under Grant No. DE-SC002567.
M.K. acknowledges support by the University of Iowa.


\begin{thebibliography}{59}
\bibitem[{not({\natexlab{a}})}]{note:11}
\bibinfo{note}{When both radiation components are circularly polarized the
  interference is absent due to the angular momentum selection rule.}
\bibitem[{not({\natexlab{e}})}]{note:0}
\bibinfo{note}{Here, we limit ourselves to processes involving a single
  $\omega_2 = N\omega$ photon.}
\bibitem[{not({\natexlab{f}})}]{note:16}
\bibinfo{note}{In addition, the initial and final angles $\varphi$ and $\varphi'$
  should enter the scattering amplitude with the same coefficients, because of
  the symmetry under the time reversal, $\varphi \leftrightarrow \varphi'$,
  $\omega_c \rightarrow - \omega_c$ and $\sigma \rightarrow - \sigma$.}
\bibitem[{not({\natexlab{g}})}]{note:6}
\bibinfo{note}{In the monochromatic case, the amplitude $A$ contains
  $\exp\left[ i \sigma ( \varphi + \varphi') / 2 \right]$ and the collision
  term, proportional to $\langle |A|^2 g_{\varphi -\varphi'}
  \rangle_{\varphi\varphi'}$ is angular independent resulting in a nonzero
  contribution.}
\bibitem[{not({\natexlab{i}})}]{note:8}
\bibinfo{note}{This result follows from the picture of the guiding centers
  jumping by a distance $2 R_c( \sin \varphi - \sin \varphi')$ across the Hall
  field.\citep{vavilov:2004,vavilov:2007}}.
\bibitem[{not({\natexlab{j}})}]{note:15}
\bibinfo{note}{If $\sigma_1=-\sigma_2$ one again has a trivial case of
  $N_1=N_2$.}
\bibitem[{not({\natexlab{k}})}]{note:1}
\bibinfo{note}{In a typical high mobility two-dimensional electron system, $\ttr/\tq \gg 1$ and the domain
  of magnetic fields where $\ttr^{-1}\ll\oc\lesssim\tau^{-1}_q$ is quite wide.}
\bibitem[{not({\natexlab{m}})}]{note:9}
\bibinfo{note}{The amplitude of $a_n\mathcal{F}_n/2\pi\eac$ is determined by
  $_{3}C_n \cdot n=$ 3, 6, and 3 for $n =$ 1, 2, and 3, respectively.}
\end{thebibliography}
\end{document}